\documentclass[11pt]{article}
\usepackage{graphicx}
\topmargin - 2cm
\oddsidemargin - 0.1cm
\evensidemargin - 5cm

\begin{document}

\title{\small{{\bf A FRACTIONAL ANALOG OF THE DUHAMEL'S PRINCIPLE }}}
\author{\bf Sabir Umarov$^{1,2}$, Erkin Saydamatov$^{1}$}  
\date{}
\maketitle
\begin{center}

$^{1}$ {\it Department of Mathematics and Mechanics\\ 
The National University of Uzbekistan, 
Tashkent, Uzbekistan}\\

$^{2}$ {\it Department of Mathematics and Statistics\\ 
University of New Mexico, 
Albuquerque, NM 87131, USA}\\
\end{center}
\begin{abstract}
The well known Duhamel's principle allows to reduce the Cauchy 
problem for a linear inhomogeneous partial differential equation
to the Cauchy problem for the corresponding homogeneous equation.
In the paper one of the possible generalizations of the 
classical Duhamel's principle to the time-fractional 
pseudo-differential equations is established. 
\end{abstract}

\section{Introduction}

The role of the classical "Duhamel's principle", introduced by Jean-Marie-Constant Duhamel 
in 1830th, is well known. The main idea of this famous 
principle is to reduce the Cauchy problem for a given linear inhomogeneous partial differential
equation to the Cauchy problem for the corresponding homogeneous equation,
which is more simpler to handle.
In this paper we establish a fractional analog of the Duhamel's principle
with respect to the following Cauchy problem for inhomogeneous
time-fractional pseudo-differential equations 
\begin{equation}
\label{equation}
D_*^\alpha u(t,x)=A(D_x)u(t,x)+f(t,x),\quad t>0,\quad x\in R^n,
\end{equation}
\begin{equation}
\label{cauchy}
\frac{\partial^{k}u}{\partial t^{k}}(0,x)= \varphi_{k}(x),\quad
x\in R^n,\,\,k = 0,..., m - 1, 
\end{equation}
where  
$\alpha \in (m-1, m],$ $m \geq 1$ is an integer;
$f(t,x)$ and $\varphi_{k}(x),\,\,k = 0,...,\,m - 1,$ are
given functions in certain spaces defined later; \quad $D_x =
(D_1,...,D_n) \quad D_j = -i\frac{\partial}{\partial x_j}, \quad
j= 1,...,n; \quad A(D_x)$ is a pseudo-differential operator with a
symbol $A(\xi)$ defined in an
open domain $G\subseteq R^n;$  and \,$D_*^\alpha$\, is the
operator of fractional differentiation of order $\alpha$ in the
Caputo sense (see, for example, \cite{Caputo,GorenfloMainardi1,GLU})
\begin{equation}
\label{Caputo}
  ({D}_{\ast}^{\alpha}) f(t) = 
  \cases{
    f(t),& $\alpha=0$,\cr
    { 
      \frac{1}{\Gamma(1-\alpha)}
      \int_0^{t} \frac{f^{'}(\tau) d \tau}{(t-\tau)^{\alpha}}
    }, & $0<\alpha<1$,\cr
    { 
      D_{\ast}^{\alpha -m} \left( \frac{d^m}{dt^m} f(y) \right)
    },& $m\le \alpha <m+1,\ m=1,2,\dots$.
  }
\end{equation}
Note that at the same time other generalizations of the Duhamel's principle based on other 
definitions of fractional derivatives are possible, but in this paper
we do not stop by these possibilities.

In our analysis we essentially relay on the results obtained in the paper
\cite{GLU}, where the Cauchy problem for fractional order homogeneous
pseudo-differential equations of arbitrary order $\alpha$  ($\alpha > 0$)
is studied. Note that fractional order inhomogeneous 
equations can not be directly
reduced to the corresponding homogeneous equations and the classical Duhamel's principle is not
applicable. To our best knowledge, solution of the Cauchy
problem for general inhomogeneous fractional order pseudo-differential equations 
require construction and estimation of the
corresponding Green's function or combination of classic Duhamels principle and some integral equations 
\cite{Bajlekova,BaeumerKuritaMeerschaert,Ahn,Podlubni,Saydamatov2006}.
The fractional Duhamel's principle established in the current paper can be applied directly to inhomogeneous fractional order equations
reducing them to corresponding homogeneous equations, at least, in the
framework of the Cauchy problem (\ref{equation}), (\ref{cauchy}).

This paper is organized as follows. In Section 2 we recall some facts related to the fractional Caputo derivative, the classic Duhamel's principle, and
the pseudo-differential operators with singular symbols. In Section 3 we formulate the main result of the present paper, namely
the fractional analog of the Duhamel's principle. In this section we also demonstrate some applications
of the obtained principle.

\section{Preliminaries}

\subsection{On the Caputo fractional derivative} 
It follows from the definition (\ref{Caputo}) of the Caputo fractional 
derivative of order $\alpha \in (m-1,m]$ for a given function $f(t)$, that its $m$-th derivative $f^{(m)}(t)$ has to exist and satisfy certain integrability 
conditions. We require throughout the paper that these conditions are fulfilled. 
We note also that if $\alpha=m$, where $m \ge 1$ is an integer number, then $D_{\ast}^{\alpha}f(t)=f^{m}(t),$ which immediately follows from the definition. 

Further, denote by
$J^{\gamma}, \, \gamma \ge 0,$ the fractional order integration operator  
\[
J^{\gamma} f(t)=\frac{1}{\Gamma (\gamma)} \int_0^t (t-\tau)^{\gamma - 1} f(\tau) d \tau,
\]
with $J^{0} = I,$ $I$ is the identity operator. It is well known \cite{GorenfloMainardi1}, 
that the family $\{J^{\gamma}, \, \gamma \ge0\}$ possesses the semi-group
property. Namely, $J^{\gamma_1 + \gamma_2} = J^{\gamma_1} J^{\gamma_2} = J^{\gamma_2} J^{\gamma_1}, \, \gamma_1 \ge 0, \, \gamma_2 \ge 0.$
The Caputo derivative can be written in the form (see \cite{GorenfloMainardi1})
\begin{equation}
\label{Caputo1}
D_{\ast}^{\alpha} f(t) = J^{m-\alpha} f^{(m)}(t), \, m-1 < \alpha \le m.
\end{equation}
If a function $f(t)$ satisfies the condition
\begin{equation}
\label{zero}
f^{(k)}(0)=0, \, k=0,..., m-1,
\end{equation}
then for $\alpha \in (m-1, m]$ and a real $\beta \ge 0$ we have
\begin{equation} 
\label{relation}
J^{\beta + \alpha}f(t) = J^{\beta + m} D_{\ast}^{m-\alpha} f(t).
\end{equation}
The relation (\ref{relation}) is obviously fulfilled, if $\alpha = m.$ Assume $m-1 <\alpha <m.$
Then for $f(t),$ satisfying (\ref{zero}) the equality $f(t) = J^m f^{(m)}(t)$ holds. Hence,
\begin{equation}
\label{relation1}
J^{\beta + \alpha} f(t) = J^{\beta + \alpha} J^m f^{(m)}(t)=J^{\beta + m} J^{\alpha} f^{(m)}(t).
\end{equation}
From the other hand $D_{\ast}^{m-\alpha}f(t)=J^{m-(m-\alpha)}f^{(m)}(t)=J^{\alpha}f^{(m)}(t).$ This fact and 
the relation (\ref{relation1}) imply (\ref{relation}).

Further, if $f$ does not satisfy the conditions (\ref{zero}), then still the relation analogous to (\ref{relation}) holds,
but with $D_+^{m-\alpha}$, the fractional derivative in the sense of Riemann-Liouville, instead of $D_{\ast}^{m-\alpha}.$
Namely, in this case the relation
\begin{equation} 
\label{relation2}
J^{\beta + \alpha}f(t) = J^{\beta + m} D_{+}^{m-\alpha} f(t),
\end{equation}
holds, where
\[
D_{+}^{\gamma} f(t) = \frac{1}{\Gamma(m-\gamma)}\frac{d^{m}}{dt^m}\int_0^t \frac {f(\tau)}{(t-\tau)^{\gamma +1 - m} d\tau}, \, m-1 <\gamma <m,
\]
and $D_+^{m} f(t) = f^{m}(t).$
Indeed, in this case we have $f(t) = J^m f^{(m)}(t) + \sum_{k=0}^{m-1}\frac{f^{(k)}(0)}{k!}t^k$, which yields
\[
J^{\beta+\alpha}f(t) = J^{\beta + \alpha} \left(J^m f^{(m)}(t) + \sum_{k=0}^{m-1} \frac{f^{(k)}(0)}{k!}t^k \right)
\]
\[
= J^{\beta+m} J^{\alpha}f^{(m)}(t) + \sum_{k=0}^{m-1} \frac{f^{(k)}(0)t^{k+\beta+\alpha}}{\Gamma(k + \alpha + \beta + 1)}.
\] 
From the other hand using the relation (see \cite{GorenfloMainardi1})
\[
D_+^{\alpha}f(t) = D_{\ast}^{\alpha}f(t) + \sum_{k=0}^{m-1}\frac{f^{(k)}(0)}{\Gamma(k-\alpha+1)}t^{k-\alpha}, \, m-1 < \alpha <m,
\]
we have
\[
J^{\beta+m}D_+^{m-\alpha}f(t) = J^{\beta + m}J^{\alpha}f^{(m)}(t) + \sum_{k=0}^{m-1} \frac{f^{(k)}(0)t^{k+\beta+\alpha}}{\Gamma(k + \alpha + \beta + 1)}.
\]

\subsection{Basic spaces of functions and pseudo-differential operators}
In this section we briefly recall some necessary notions and facts,
which we use in this paper referring the reader to \cite{GLU,Umarov}, for details.
Let $G$ be an open domain in $R^n$ and the system of open sets
$\{g_k\}_{k=0}^{\infty}$ be a locally finite covering of  $G$,
i.e.
$$G = \bigcup_{k=0}^{\infty} g_k ,\quad g_{k}\subset\subset G.$$
Let any compact set $K \subset G$ has a nonempty intersection with
finitely many sets $g_k.$ Denote by $\{\theta_k\}_{k=0}^{\infty}$
a smooth partition of unity of $G$.
Further,  let $1 < p < \infty$ and a function $f(x)$ be in
$L_{p}(R^n)$ whose Fourier transform $F\,f$ has a compact support in $G$. 
The set of all such functions
endowed with the convergence defined in Definition 1 is denoted
by $\Psi_{G,p}(R^n)$.

\vspace{.3cm}

{\bf Definition 1.}{\it A sequence of functions $\,f_{m}\in
\Psi_{G,p}(R^n),\,\,m = 1,\,2,\,3,...$ is said to converge to an
element $f_{0}\in \Psi_{G,p}(R^n)$ iff:

1) there exists a compact set $K \subset G$ such that the support
$supp\,Ff_{m}\subset K$ \,\,for all\,\, $m \in N;$

2) the norm $\|f_{m} - f_{0}\|_{L_{p}} = (\,\int_{R^{n}}|f_{m}(x)
- f_{0}(x)|^{p}dx\,)^{\frac{1}{p}}\,\rightarrow\,0$\,\, for\,\, $m
\rightarrow\,\infty.$
}
\vspace{.3cm}

According to the Paley-Wiener-Schwartz theorem elements of
$\Psi_{G,p}(R^n)$ are entire functions of exponential type, which,
restricted to $R^n$, are in the space $L_{p}(R^n).$

The space $\Psi_{G,p}(R^n)$ can be represented as an inductive
limit of some Banach spaces. Namely, let
$$G_{N} = \bigcup _{k=1}^{N}g_{k},\quad \chi_{_{N}}(\xi) = \sum _{k=1}^{N}\theta_{k}(\xi).$$
Denote by $\Psi_{N}$ the set of functions $f \in L_{p}(R^n)$
satisfying the following conditions:

a) $supp\,Ff \subset G_{N} ;$

b) $supp\,Ff \cap supp \,\,\theta_{j} = \emptyset$ \, \, for \,\,
$j
> N ;$

c) $p_{_{N}}(f) = \|F^{-1}\chi_{_{N}}Ff\|_{L_{p}} < \infty.$\\
Here by $F^{-1}$ we denote the operator inverse to the Fourier
transform $F$. It is not hard to verify that (see \cite{GLU,Umarov})
$$\Psi_{G,p} = ind\,\lim_{N\rightarrow \infty} \Psi_{N}.$$

Let $A(D)$ be a pseudo-differential operator with a symbol
$A(\xi)$, which is a real-analytic\footnote{This condition can be essentially weakened. 
See the construction in \cite{UmarovGorenflo,UmarovSteinberg}, where continuous 
symbols are considered.} function in $G$. Outside of $G$
or on its boundary $A(\xi)$ may have singularities of arbitrary
type. For a function $\varphi(x) \in \Psi_{G,p}(R^n)$ the operator
$A(D)$ is defined by the formula
$$A(D)\,\varphi(x) = \frac{1}{(2\pi)^n}\int_{R^n}A(\xi)F\varphi(\xi)e^{ix\xi}d\xi =
\frac{1}{(2\pi)^n}\int_{G}A(\xi)F\varphi(\xi)e^{ix\xi}d\xi.$$

As is shown in \cite{Umarov,GLU}  the space $\Psi_{G,p}(R^n)$ is 
invariant with respect to the action of such pseudo-differential
operators and these operators act continuously.

\vspace{0.3cm}
\subsection{The classic Duhamel's principle}
Recall the classic Duhamel's integral and Duhamel's principle. 
The Duhamel's integral (see, e.g. \cite{TyhonovSamarski,BersJohnSchechter}) is used for 
representation of a solution of the Cauchy problem for a given inhomogeneous
linear partial differential equation with homogeneous initial conditions via the solution
of the Cauchy problem for the corresponding homogeneous equation. Consider
the Cauchy problem for the second order inhomogeneous differential equation
\begin{equation}
\label{e1}
\frac{\partial ^{2}u}{\partial t^{2}}(t,x) = L \, u(t,x) + f(t,x),
\quad t > 0,\,\,x \in R^{n},
\end{equation}
with homogeneous initial conditions
\begin{equation}
\label{cc1}
u(0,x) = 0,\quad \frac{\partial u}{\partial t}(0,x) = 0, 
\end{equation}
where \,$L$\, is a linear differential operator with coefficients
not depending on \,$t$,\, and containing the temporal derivatives of 
order, not higher then $1$. 
Further, let a sufficiently smooth function \,$v(t,
\tau, x),\,\,t \geq \tau,\,\,\tau \geq 0,\,\, x \in R^{n}$,\, be
for $t > \tau$ a solution of the homogeneous equation
$$\frac{\partial^{2}v}{\partial t^{2}}(t, \tau, x) = L \, v(t, \tau, x),$$
satisfying the following conditions:
$$v(t, \tau, x)|_{t=\tau} = 0,\quad \frac{\partial v}{\partial t}(t, \tau, x)|_{t=\tau} = f(\tau, x).$$
Then a solution of the Cauchy problem (\ref{e1}), (\ref{cc1}) is given by means
of the Duhamel's integral
$$u(t,x) = \int_{0}^{t}v(t, \tau, x)d\tau.$$
The formulated statement has been entitled as the "Duhamel's principle".

An analogous construction is possible in the case of the Cauchy
problem with a homogeneous initial condition for the first order inhomogeneous partial differential equation
$$\frac{\partial u}{\partial t}(t,x) = M \, u(t,x) + f(t,x), \quad t > 0, \,\,x \in R^{n},$$
where \,$M$\, is a linear differential operator containing only spatial derivatives, and with coefficients
not depending on $t$.

\vspace{.3cm}
\subsection{The Duhamel's principle for integer $\alpha = m, \, m \ge 1$} Consider the Cauchy problem (1), (2) in the case of integer
\,\,$\alpha = m \geq 1, \, D_{*}^{\alpha} \equiv
\frac{\partial^m}{\partial t^m}$, i.e.
\[
\frac{\partial^{m} u}{\partial t^m}(t,x) =
 A(D_x)u(t,x) + f(t,x),\quad t > 0,\,\,x \in R^n, 
\]
\[
\frac{\partial^{k}u}{\partial t^k}(0,x) = \varphi_{k} (x), \quad x \in R^n, \,\,k = 0,..., m-1.
\]
In this case the Duhamel's principle is formulated as follows.
Let  $U(t,\tau,x)$ be a solution of the Cauchy problem for a homogeneous equation
\begin{equation}
\label{U}
\frac{\partial^{m} U}{\partial t^m} = A(D_x)U, \quad 0 < \tau < t,
\end{equation}
\begin{equation}
\label{m-2}
\frac{\partial^{k}U}{\partial t^k}(t,\tau,x)|_{t=\tau} = 0,\,\,\,k = 0,...,m-2,
\end{equation}
\begin{equation}
\label{m-1}
\frac{\partial^{m-1}U}{\partial t^{m-1}}(t,\tau,x)|_{t=\tau} = f(\tau,x).
\end{equation}
Then the function
\begin{equation}
\label{solgen}
u(t,x) = \int_{0}^{t} U(t, \tau, x)d\tau
\end{equation}
is a solution of the Cauchy problem
\begin{equation}
\label{int}
\frac{\partial^{m} u}{\partial t^m} - A(D_x)u = f(t,x), 
\end{equation}
\begin{equation}
\label{ccgen}
\frac{\partial^{k}u}{\partial t^k}(0,x) = 0,\,\,\,k = 0,...,\, m-1.
\end{equation}

The proof of this statement can be found, for instance, in \cite{Dubinskii}. However, for the completeness, we 
reproduce the proof.

Obviously $u(0,x) = 0.$ Further, for the first order derivative 
$$
\frac{\partial u}{\partial t}(t,x) = U(t, t, x) + \int_{0}^{t}\frac{\partial}{\partial t}U(t, \tau, x)d\tau, 
$$
it follows from (\ref{m-2}) that
$\frac{\partial u}{\partial t}(0,x) = 0.$ Analogously we calculate $\frac{\partial^k u}{\partial t^k}(0,x) = 0, \, k=1,...,m-2.$
Consequently, for the derivative of $(m-1)$-th order
$$
\frac{\partial^{m-1}}{\partial t^{m-1}}u(t,x) = \frac{\partial^{m-2}}{\partial t^{m-2}}U(t, t, x) +
\int_{0}^{t}\frac{\partial^{m-1}}{\partial t^{m-1}}U(t, \tau,
x)d\tau, 
$$
we obtain
$
\quad \frac{\partial^{m-1}}{\partial t^{m-1}}u(0,x) = 0.$ 
Therefore, the function \,$u(t,x)$\, in (\ref{solgen}) satisfies  the
initial conditions (\ref{ccgen}).
Moreover,
$$\frac{\partial^{m}u}{\partial t^m} - A(D_{x})u = \frac{\partial^m}{\partial t^m}
\int_{0}^{t}U(t,\tau,x)d\tau -
A(D_{x})\int_{0}^{t}U(t,\tau,x)d\tau =$$
$$= \frac{\partial^{m-1}}{\partial t^{m-1}}U(t,t,x) +
\int_{0}^{t}\frac{\partial^m}{\partial t^m}U(t,\tau,x)d\tau -
 \int_{0}^{t}A(D_{x})U(t,\tau,x)d\tau =$$
$$= f(t,x) + \int_{0}^{t}[\frac{\partial^m}{\partial t^m}U(t,\tau,x) - A(D_{x})U(t,\tau,x)]d\tau = f(t,x).$$
Hence, $\,u(t,x)\,$ in (\ref{solgen}) satisfies the equation (\ref{int}) as well.

\vspace{0.3cm}
\subsection{The representation formula for a solution of the Cauchy problem for homogeneous fractional order equations}
Now we consider the Cauchy problem (\ref{equation}), (\ref{cauchy}) for arbitrary $\alpha > 0$. Note that in this
case Duhamel's principle can not be applied directly. For the Cauchy
problem (\ref{equation}), (\ref{cauchy}) in the homogeneous case (i.e. $f(t,x) \equiv 0$ in Equation (\ref{equation})) 
the following representation formula for a solution was obtained in \cite{GLU}:
\begin{equation}
\label{representation} 
u(t,x) = \sum_{k=1}^{m}J^{k-1}E_\alpha(t^\alpha A(D_x)) \varphi_{k-1}(x),
\end{equation}
where $J^{k}$ is the $k$-th order integral operator,
$E_\alpha(t^\alpha A(D_x))$ is a pseudo-differential
operator with the symbol $E_\alpha(t^\alpha A(\xi))$ and
\,$E_{\alpha}(z)$\, is the Mittag-Leffler function (see \cite{Djrbashian66})
$$
E_{\alpha}(z) = \sum_{k=0}^{\infty} \frac{z^k}{\Gamma(1 + \alpha k)}.
$$

\section{Main results}

\subsection{A fractional Duhamel's principle in the case $0 < \alpha <1$}
Assume $0 < \alpha < 1$. First we formulate a formal fractional analog of the Duhamel's principle and then we show how to apply this
principle in various situations. 

\vspace{.3cm}

{\bf Theorem 1.}
{\it Suppose that $V(t,\tau,x), \, 0 \le \tau \le t, \, x \in R^n,$ is a solution of the Cauchy problem for homogeneous equation 
\begin{equation}
\label{Eq11}
D_{*}^{\alpha}V(t,\tau,x) - A(D_{x})V(t,\tau,x) = 0, \quad t > \tau, \quad x \in R^{n}, 
\end{equation}
\begin{equation}
\label{Con11}
V(\tau,\tau,x) = D_{*}^{1-\alpha}f(\tau,x), \quad x \in R^n, 
\end{equation}
where $f(t,x)$ is a given function satisfying the condition $f(0,x) = 0$. Then 
\begin{equation}
\label{sol11}
v(t,x) = \int_{0}^{t}V(t,\tau,x)d\tau
\end{equation}
is a solution of the inhomogeneous Cauchy problem 
\begin{equation}
\label{inheq11}
D_{*}^{\alpha}v(t,x) - A(D_x)v(t,x) = f(t,x), 
\end{equation}
\begin{equation}
\label{homcon11}
\quad v(0,x) = 0.
\end{equation}
}

\vspace{.3cm}

{\bf Proof.}
Notice that in accordance with (\ref{representation}) a solution of the Cauchy problem (\ref{Eq11}),(\ref{Con11}) is represented in the form
\begin{equation}
\label{V}
V(t,\tau,x)=E_{\alpha}((t-\tau)^{\alpha} A(D_x) )D_{\ast}^{1-\alpha}f(\tau,x).
\end{equation}

Further, apply the operator $J^{\alpha}$ to both sides of Equation (\ref{inheq11}) and use the relation
$J^{\alpha}D_{\ast}^{\alpha}v(t,x) = v(t,x)-v(0,x),$ to obtain
\[
v(t,x) - J^{\alpha} A(D_x) v(t,x) = J^{\alpha} f(t,x).
\]
A solution of the last equation can be represented as
\[
v(t,x)= \sum_{n=0}^{\infty}J^{\alpha n + \alpha} A^n(D_x) f(t,x).
\]
It follows from (\ref{relation}) (with $\beta=\alpha n$ and $m=1$) that for arbitrary function $g(t)$ satisfying the condition $g(0)=0$, there holds
$J^{\alpha n +\alpha} g(t) = J^{\alpha n + 1} D_{\ast}^{1-\alpha}g(t).$
Taking this into account we have
\[
v(t,x)= \sum_{n=0}^{\infty}J^{\alpha n + 1} A^n(D_x) D_{\ast}^{1-\alpha} f(t,x) =
\int_0^{t} \sum_{n=0}^{\infty} \frac{(t-\tau)^{\alpha n} A^n (D_x)}{\Gamma(\alpha n + 1)}  D_{\ast}^{1-\alpha} f(t,x) =
\]
\begin{equation}
\label{v}
\int_0^{t} E_{\alpha}((t-\tau)^{\alpha} A(D_x)) D_{\ast}^{1-\alpha} f(t,x).
\end{equation}
Comparing (\ref{V}) and (\ref{v}) we arrive at (\ref{sol11}).

{\bf Remark.}
{\it 
\begin{enumerate}
\item 
If $\alpha = 1,$ then Theorem 1 coincides with the classic Duhamel's principle, because in this case $D_{*}^{1-\alpha}f(t,x) = f(t,x)$. 
\item
It is well-known that the fractional derivative $D_{*}^{1-\alpha}f(t), \, 0 < \alpha < 1$ exists, if $f(t) \in AC[0 \le t \leq T]$, 
where $T$ is a positive finite number and $AC[0,T]$ is the class of absolutely continuous functions (see \cite{SKM}). 
\item 
The condition $f(0,x) = 0$ in Theorem 1 is not essentially restrictive. 
For arbitrary  $f(t,x)$ in the formulation of Theorem  the Cauchy condition (\ref{Con11}) has to be replaced by
\[
V(\tau,\tau,x) = D_{+}^{1-\alpha}f(\tau,x), \quad x \in R^n, 
\]
where $D_{+}^{1-\alpha}$ is the operator of fractional differentiation of order $1 - \alpha$ in the Riemann-Liouville sense. 
\end{enumerate}
}

\vspace{.3cm}

Denote by $C^{(m)}[t > 0;\, \Psi_{G,2}(R^n)]$ \, and by \,$AC[t > 0;\, \Psi_{G,2}(R^n)]$ \, the space of \,\,$m$\,\,times
continuously differentiable functions and the space of absolutely continuous functions on \,$(0; +\infty)$ \, with values ranging in
the space \,$\Psi_{G,2}(R^n)$, respectively.

\vspace{.3cm}

{\bf Theorem 2.}
{\it  
Let 
$\varphi_{0}(x) \in \Psi_{G,2}(R^n),\quad f(t,x) \in AC[t \geq 0;\, \Psi_{G,2}(R^n)]$,\quad $D_{*}^{1-\alpha}f(t,x) \in C[t \geq 0;\,\Psi_{G,2}(R^n)]$
and $f(0,x) = 0.$ 
Then the Cauchy problem
(1), (2) (with \,$0 < \alpha < 1$) has a unique solution  
$
u(t,x) \in C^{(1)}[t>0;\, \Psi_{G,2}(R^n)] \, \cap \, C[t \geq 0;\, \Psi_{G,2}(R^n)].
$ 
This
solution is given by the representation
\begin{equation}
\label{repgen}
u(t,x) = E_\alpha(t^\alpha A(D_x))\varphi_{0}(x) + \int_{0}^{t}E_\alpha((t-\tau)^\alpha A(D_x))D_{*}^{1-\alpha} f(\tau,x)d\tau. 
\end{equation}
}

{\bf Proof.} 
The representation (\ref{repgen}) is a simple implication of (\ref{representation}) and Theorem 1.
The first term in (\ref{repgen}) is studied in \cite{GLU} in detail. Denote by $v(t,x)$ the second term in (\ref{repgen}).
For a fixed $t > 0$ making use of the semi-norm of
$\Psi_{N}$ we have
$$p_{_{N}}^{2}(v(t,x)) = \|F^{-1}\chi_{_{N}}Fv\|_{L_{2}}^{2} = \|\chi_{_{N}}Fv\|_{L_{2}}^{2} =$$
$$= \int_{R^n}|\chi_{_{N}}(\xi)|^{2}\cdot
|\int_{0}^{t}E_{\alpha}((t-\tau)^{\alpha}A(\xi))FD_{*}^{1-\alpha}f(\tau,\xi)d\tau|^{2}d\xi.$$
For $\chi_{_{N}}(\xi)$ there exists a compact set $K_{_{N}}
\subset G$ such that $supp \,\chi_{_{N}}(\xi) \subset K_{_{N}}$.
By using Cauchy-Bunjakowski's inequality we get the estimate
$$p_{_{N}}^{2}(v(t,x)) \leq \int_{K_{_{N}}} |\chi_{_{N}}(\xi)|^{2}\cdot
\int_{0}^{t}|E_{\alpha}((t-\tau)^{\alpha}A(\xi))|^{2}d\tau \cdot
\int_{0}^{t}|FD_{*}^{1-\alpha}f(\tau,\xi)|^{2}d\tau d\xi.$$ The
function\,\,
$\int_{0}^{t}|E_{\alpha}((t-\tau)^{\alpha}A(\xi))|^{2}d\tau$
\,\,is bounded on $K_{_{N}}$. Consequently, there exists a
constant $C_{_{N}} > 0$ such that
$$p_{_{N}}^{2}(v(t,x)) \leq C_{_{N}} \int_{K_{_{N}}} |\chi_{_{N}}(\xi)|^{2}\cdot
\int_{0}^{t}|FD_{*}^{1-\alpha}f(\tau,\xi)|^{2}d\tau d\xi \leq $$
$$
\leq C_{_{N}} \int_{0}^{t}\int_{R^n} |\chi_{_{N}}(\xi)|^{2}\cdot |FD_{*}^{1-\alpha}f(\tau,\xi)|^{2}d\xi d\tau =$$ $$= C_{_{N}}
           \int_{0}^{t}\|\chi_{_{N}}(\xi)FD_{*}^{1-\alpha}f(\tau,\xi)\|_{L_{2}}^{2} d\tau =
                                          C_{_{N}}\int_{0}^{t}p_{_{N}}^{2}(D_{*}^{1-\alpha}f(\tau,x))d\tau.
$$
It follows from the condition  \,$D_{*}^{1-\alpha}f(t,x)\in C[t
\geq 0;\,\Psi_{G,2}(R^n)]$\, that the function\,
$p_{_{N}}(D_{*}^{1-\alpha}f(\tau,x))$ \,is continuous with respect
to\, $\tau \in (0;t)$ \,and for a fixed \,$t > 0$ \, and some $N_1$ the estimate
$$
p_{_{N}}^{2}(v(t,x)) \leq C_{_{N}}\cdot t \cdot \sup_{0<\tau<t}p_{_{N}}^{2}(D_{*}^{1-\alpha}f(\tau,x))  
                                   \leq C_{_{N_1}}\cdot t \cdot \sup_{0<\tau<t}p_{_{N_1}}^{2}(f(\tau,x)) 
$$ 
holds. Consequently, for every fixed \,$t \in
(0;+\infty)$\, the function $\,v(t,x)\,$ in (\ref{sol11}) belongs to the
space $\Psi_{G,2}(R^n)$. The analogous estimate is valid for
\,$\frac{\partial}{\partial t}v(t,x).$ Thus 
$v(t,x) \in C^{(1)}[t > 0;\,\Psi_{G,2}(R^n)]\,\cap \,C[t \geq 0;\,\Psi_{G,2}(R^n)]$. 
Hence, 
$v(t,x) \in C^{(1)}[t > 0;\,\Psi_{G,2}(R^n)]\,\cap \,C[t \geq 0;\,\Psi_{G,2}(R^n)]$, 
as well.
The uniqueness of a solution follows from the representation formula for a solution of the homogeneous Cauchy problem.

\subsection{A fractional Duhamel's principle in the case  of arbitrary $ \alpha >0 $}
Now we consider the Cauchy problem (1), (2) for arbitrary order
\,\,$\alpha,\quad m - 1 < \alpha \le m \in N$. Obviously, in this case $0 < m - \alpha \le 1$.

\vspace{.3cm}

{\bf Theorem 3.}
{\it
Assume $m \ge 1, \, m-1 < \alpha \le m$, and $V(t,\tau, x)$ is a solution of the Cauchy problem for  
the homogeneous equation (\ref{Eq11})
with the Cauchy conditions
\begin{equation}
\label{CC121}
\frac{\partial ^{k}V}{\partial t^{k}}(t,\tau,x)|_{t=\tau} = 0,\,\,\,k = 0, ..., m-2,
\end{equation}
\begin{equation}
\label{CC122}
\frac{\partial ^{m-1}V}{\partial t^{m-1}}(t,\tau,x)|_{t=\tau} = D_{*}^{m-\alpha}f(\tau,x),
\end{equation}
where $f(t,x), \, t>0,\, x\in R^n$, is a given function satisfying the conditions  $\frac{\partial^k f(0,x)}{\partial t^k} = 0, \, k=0,...,m-1$. 
Then 
$$
v(t,x) = \int_{0}^{t}V(t,\tau,x)d\tau \eqno{(\ref{sol11})}
$$
is a solution of the Cauchy problem for the inhomogeneous equation (\ref{inheq11}) with the homogeneous Cauchy conditions
\begin{equation}
\label{CC123}
\frac{\partial ^{k}v}{\partial t^{k}}(0,x) = 0,\,\,\,k = 0, ..., m-1.
\end{equation}
}
\vspace{.3cm}

{\bf Proof.}
It follows from the representation formula (\ref{representation}) that 
\begin{equation}
\label{V1}
V(t,\tau,x) = J^{m-1} E_{\alpha}((t-\tau)^{\alpha}A(D_x)) D_{\ast}^{m - \alpha} f(\tau,x)
\end{equation}
solves the Cauchy problem for Eq. (\ref{Eq11}) with the initial conditions (\ref{CC121}), (\ref{CC122}). 

Further, apply the operator $J^{\alpha}$ to both sides of the equation (\ref{inheq11}) and obtain
\begin{equation}
\label{step2}
v(t,x) - \sum_{j=0}^{m-1} \frac{t^j v^{j}(0,x)}{j!} - J^{\alpha} A(D_x) v(t,x) = J^{\alpha} f(t,x).
\end{equation}
Taking into account the conditions (\ref{CC123}), we rewrite Eq. (\ref{step2})  in the form 
\[
v(t,x) -  J^{\alpha} A(D_x) v(t,x) = J^{\alpha} f(t,x).
\]
A solution of this equation is represented as
\[
v(t,x)= \sum_{n=0}^{\infty}J^{\alpha n + \alpha} A^n(D_x) f(t,x).
\]
It follows from (\ref{relation}) (with $\beta = \alpha n$) that for arbitrary function $g(t)$ satisfying the conditions $g(0)=0, \, g^{'}(0)=0, ... , g^{(m-1)}(0)=0$, there holds
$J^{\alpha n +\alpha} g(t) = J^{\alpha n + 1} J^{m-1} (D_{\ast}^{m-\alpha}g(t)).$
Taking this into account, we have
\[
v(t,x)= \sum_{n=0}^{\infty}J^{\alpha n + 1} J^{m-1} A^n(D_x) D_{\ast}^{m-\alpha} f(t,x) =
\]
\[
\int_0^{t} J^{m-1} \sum_{n=0}^{\infty} \frac{(t-\tau)^{\alpha n} A^n (D_x)}{\Gamma(\alpha n + 1)}  D_{\ast}^{m-\alpha} f(t,x) =
\]
\begin{equation}
\label{v1}
\int_0^{t} J^{m-1} E_{\alpha}((t-\tau)^{\alpha} A(D_x)) D_{\ast}^{m-\alpha} f(t,x).
\end{equation}
Comparing (\ref{V1}) and (\ref{v1}) we obtain (\ref{sol11}), and hence, the proof of the theorem.

\vspace{.3cm}

{\bf Remark.} 
{\it Note that if $\alpha = m,$ then Theorem 3 recovers the known classic Duhamel's principle we mentioned above (see, subsection 2.4). 
Moreover, the conditions $\frac{\partial^k f(0,x)}{\partial t^k} = 0, \, k=0,...,m-1$, which we required in Theorem, are not essential. 
For arbitrary  $f(t,x)$, as a consequence of the relationship (\ref{relation2}), the formulation of the fractional Duhamel's principle 
takes the following form.
}
\vspace{.3cm}

{\bf Theorem 4.}
{\it
Assume $m \ge 1, \, m-1 < \alpha \le m$, and $V(t,\tau, x)$ is a solution of the Cauchy problem for  
the homogeneous equation (\ref{Eq11}) 
with the Cauchy conditions
\begin{equation}
\label{CC131}
\frac{\partial ^{k}V}{\partial t^{k}}(t,\tau,x)|_{t=\tau} = 0,\,\,\,k = 0, ..., m-2,
\end{equation}
\begin{equation}
\label{CC132}
\frac{\partial ^{m-1}V}{\partial t^{m-1}}(t,\tau,x)|_{t=\tau} = D_{+}^{m-\alpha}f(\tau,x),
\end{equation}
where $f(t,x), \, t>0,\, x\in R^n$, is a given function. 
Then $v(t,x)$ defined in (\ref{sol11})
is a solution of the following Cauchy problem for the inhomogeneous equation 
\[
D_{*}^{\alpha}v(t,x) - A(D_x)v(t,x) = f(t,x), 
\]
\[
\frac{\partial ^{k}v}{\partial t^{k}}(0,x) = 0,\,\,\,k = 0, ..., m-1.
\]
}
\vspace{.3cm}

Theorem 3 and Theorem 4 allow to generalize the result of the paper \cite{Saydamatov2006} for arbitrary $\alpha > 0.$

\vspace{.3cm}

{\bf Definition 2.}
{\it A function   
$
u(t,x) \in C^{(m)}[t>0;\, \Psi_{G,2}(R^n)] \, \cap \, C^{(m-1)}[t \geq 0;\, \Psi_{G,2}(R^n)]
$ 
is called a solution of the problem (1), (2) if
it satisfies the equation (1) and the initial conditions (2)
pointwise.
}

\vspace{.3cm}

{\bf Theorem 5.}
{\it 
Let 
$\varphi_{k}(x) \in \Psi_{G,2}(R^n),\,\,k=0,...,\,\,m-1, \quad f(t,x) \in AC[t \geq 0;\, 
\Psi_{G,2}(R^n)]$,\quad $D_{*}^{m-\alpha}f(t,x) \in C[t \geq 0;\, \Psi_{G,2}(R^n)]$\, 
and $\frac{\partial^k f(0,x)}{\partial t^k} = 0, \, k=0,...,m-1$. 
Then the Cauchy
problem (1), (2) has a unique solution. This solution is given by
the representation
$$
u(t,x) = \sum_{k=1}^{m} J^{k-1} E_\alpha(t^\alpha A(D_x))\varphi_{k-1}(x) +
$$
\begin{equation}
\label{R}
\int_{0}^{t} J^{m-1} E_\alpha((t-\tau)^\alpha A(D_x))D_{*}^{m-\alpha} f(\tau,x)d\tau. 
\end{equation}
}
\vspace{.3cm}

{\bf Proof.} 
Splitting the Cauchy problem (\ref{equation}),(\ref{cauchy}) into the Cauchy problem for the equation (\ref{equation})
with the homogeneous initial conditions and the Cauchy problem for the homogeneous equation corresponding to (\ref{equation}) with 
the initial conditions (\ref{cauchy}), and applying Theorem 3 and representation formula (\ref{representation}), we obtain (\ref{R}).  
The fact that 
\[
\sum_{k=1}^{m} J^{k-1} E_\alpha(t^\alpha A(D_x))\varphi_{k-1}(x) \in C^{(m)}[t>0; \Psi_{G,2}(R^n)] \cap C^{(m-1)}[t \geq 0; \Psi_{G,2}(R^n)]
\]
is proved in \cite{GLU}. Further, since the $m-1$-th derivative with respect to $t$ of the last term in (\ref{R}) belongs to 
$AC[[0,T];\Psi_{G,2}(R^n)]$\footnote{$T$ is an arbitrary positive finite number.},
then the estimation obtained in the proof of Theorem 2 holds in this case as well.

\vspace{.3cm}

{\bf Remark.} 
{\it 
If $f(t,x)$ and its derivatives up to order $m-1$ with respect to $t$ do not vanish at $t=0$, then in accordance with Theorem 4 the representation formula (\ref{R}) takes the form  
\[
 u(t,x) = \sum_{k=1}^{m} J^{k-1} E_\alpha(t^\alpha A(D_x))\varphi_{k-1}(x) +
\]
\[
\int_{0}^{t} J^{m-1} E_\alpha((t-\tau)^\alpha A(D_x))D_{+}^{m-\alpha} f(\tau,x)d\tau. 
\]
}

\subsection{Examples}
{\bf 1.} Let \,$0 < \alpha < 1$ and $f(t,x)$ be a given function, $f(0,x)=0$. Consider the Cauchy problem
$$
D_*^\alpha u(t,x) = \Delta u(t,x)+ f(t,x), \, t > 0, \, x\in R^n, 
$$
$$
u(0,x) = \varphi_{0}(x). 
$$
In accordance with the fractional Duhamel's principle (Theorem 1) the influence of the external force $f(t,x)$ to the output can be count as
$$
D_*^\alpha V(\tau, t,x) = \Delta V(\tau, t,x), \, t > \tau, \, x\in R^n, 
$$
$$
V(\tau, \tau, x) = D_{\ast}^{1-\alpha} f(\tau, x). 
$$
The function $V(\tau, t, x) = E_{\alpha} ((t-\tau)^{\alpha} \Delta) D_{\ast}^{1-\alpha} f(\tau, x)$ solves this problem.
Hence, 
$$u(t,x)\, = \, E_{\alpha}(t^{\alpha} \triangle)\varphi_{0}(x)\,
+\,\int_{0}^{t}E_{\alpha}((t-\tau)^{\alpha}
\triangle)D_{*}^{1-\alpha}f(\tau,x)d\tau.$$

{\bf 2.} Similarly, if $1 < \alpha < 2$, and $F(t,x)$ is a given function, which describes the outer force, then we deal with the Cauchy problem
$$
D_*^\alpha u(t,x) = \Delta u(t,x)+ F(t,x),\quad t > 0,\quad x\in R^n, 
$$
$$
u(0,x) = \varphi_{0}(x), \quad \quad u_{t}(0,x) = \varphi_{1}(x).
$$
Again in accordance with the fractional Duhamel's principle (Theorem 4) the influence of the external force $F(t,x)$ to the output can be count as
$$
D_*^\alpha V(\tau, t,x) = \Delta V(\tau, t,x), \, t > \tau, \, x\in R^n, 
$$
$$
V(\tau, \tau,x)=0, \, \, \frac{\partial V}{\partial t}(\tau, \tau, x) = D_{+}^{2-\alpha} F(\tau, x). 
$$
The function $V(\tau, t, x) = J E_{\alpha} ((t-\tau)^{\alpha} \Delta) D_{+}^{2-\alpha} F(\tau, x)$ solves this problem. Hence,

$$u(t,x) = E_{\alpha}(t^{\alpha} \Delta)\varphi_{0}(x) +JE_{\alpha}(t^{\alpha} \Delta) \varphi_{1}(x) + $$
$$\int_{0}^{t}JE_{\alpha}((t-\tau)^{\alpha} \Delta)D_{+}^{2-\alpha}F(\tau,x)d\tau.$$


\end{document}